\documentclass[english,superscriptaddress]{revtex4-2}
\usepackage[latin9]{inputenc}
\usepackage{textcomp}
\usepackage{amsmath}
\usepackage{graphicx}
\usepackage{esint}

\makeatletter

\DeclareTextSymbolDefault{\textquotedbl}{T1}

\makeatother

\usepackage{babel}
\begin{document}
\title{Measurement of classical entanglement using interference fringes}
\author{Ziyang You}
\affiliation{Institute of Applied Physics and Materials Engineering, University
of Macau, Macau S.A.R., China}
\author{Yanxiang Wang}
\affiliation{Institute of Applied Physics and Materials Engineering, University
of Macau, Macau S.A.R., China}
\author{Zikang Tang}
\affiliation{Institute of Applied Physics and Materials Engineering, University
of Macau, Macau S.A.R., China}
\author{Hou Ian}
\email{houian@um.edu.mo}

\affiliation{Institute of Applied Physics and Materials Engineering, University
of Macau, Macau S.A.R., China}
\begin{abstract}
Classical entanglement refers to non-separable correlations between
the polarization direction and the polarization amplitude of a light
field. The degree of entanglement is quantified by the Schmidt number,
taking the value of unity for a separable state and two for a maximally
entangled state. We propose two detection methods to determine this
number based on the distinguishable patterns of interference between
four light sources derived from the unknown laser beam to be detected.
The second method being a modification of the first one has the interference
fringes with discernable angles uniquely related to the entangled
state. The maximally entangled state corresponds to fringes symmetric
about the diagonal axis at either $45^{\circ}$ or {\normalsize{}$135^{\circ}$}
direction, while the separable state corresponds to fringes symmetric
either about the $X$- or $Y$-axis or both simultaneously. States
with Schmidt number between unity and two have fringes of symmetric
angles between these two extremes. The detection methods would be
beneficial in both computation and communication applications of the
classically entangled states.
\end{abstract}
\maketitle

\section{Introduction}

Schrödinger introduced the concept of entanglement to describe non-separable
correlations among different quantum systems~\citep{Sch}, in response
to the Einstein, Podolsky, and Rosen (EPR) argument on the incompatibility
between quantum mechanics and local realism~\citep{Eist}. Later,
John Bell derived an inequality that can experimentally confirm the
nonlocality of quantum mechanics to settle the EPR argument~\citep{bell}.
Since then, various experiments have clearly shown that entangled
quantum systems can violate the Bell inequality, where Clauser, Horne,
Shimony, and Holt (CHSH) inequality is the most well-known example~\citep{Clauser}.
Apart from Bell-CHSH inequality, another measurement of entangled
states relies on the Schmidt analysis, which includes calculation
of the Schmidt number $K$ to represent the degree of entanglement~\citep{eberly_schmidt}.

In terms of nonlocality, entanglement has been viewed as a unique
feature in quantum mechanics~\citep{Paneru}. However, non-separable
correlations between different degrees of freedom in classical light
fields, termed \textquotedblleft classical entanglement,\textquotedblright{}
have been proved to exist~\citep{Lee,Spreeuw,Simon,Qian}. Several
pairs of degrees of freedom localized within a beam are correlated
in a way analogous to quantum entanglement~\citep{De zela,Aiello,Forbes},
such as that between polarization and spatial parity, between polarization
and temporal amplitude, and between polarization and path. Moreover,
recent experiments have supported the violation of Bell-CHSH inequality
among the classical degrees of freedom measured under classical entanglement,
and their correlation levels reach as high as those obtained by quantum
entanglement~\citep{Kagalwala,Qian_2,SUN,Gonzales}. Such a statistical
and physical similarity has confirmed that classical entanglement
can provide some computation and communication applications previously
supported by quantum entanglement~\citep{Holleczek}. For example,
some works have implemented quantum algorithms with classical entanglement,
like the Deutsch algorithm~\citep{Perez-Garcia}, quantum walk~\citep{Goyal},
and the quantum Fourier transform~\citep{Song}. Moreover, some works
have also demonstrated the classical entanglement applications in
communication, such as encoding~\citep{Li}, quantum channel tomography~\citep{Ndagano},
and teleportation~\citep{Guzman=002010Silva,Silva_b}. 

Though the relevant studies, including those mentioned above, are
abundant and substantial, quantifying classical entanglement in an
experimentally accessible way is less explored. Here, we propose methods
of interferometry to visualize the quantifiable degree of classical
entanglement of a beam in its projected fringe pattern. In particular,
we use Schmidt analysis~\citep{eberly_schmidt,Qian,Qian_2} to decide
this degree between the polarization direction and the polarization
amplitude of a laser beam~\citep{Qian} and design two experimental
setups --``phase method'' and ``amplitude method''-- to eventually
reveal the Schmidt number in the specific symmetry axis of the fringe
pattern. Being an improved version over the phase method on the optical
path adopted, the amplitude method would show a more apparent fringe
pattern. For instance, using the amplitude method, the fringes are
symmetric exactly about the diagonal axis at $45^{\circ}$ or $135^{\circ}$
direction for the maximal entanglement case. At the other extreme
for separable states, the symmetry axes are either on the $X$-axis,
the $Y$-axis, or both simultaneously. The classical optical setup
extends the entanglement analysis into the classical domain, which
can be beneficial in computation and communication applications of
the classically entangled states. For instance, it would open a new
path to the optics realization of quantum algorithms through classical
interferometry. In the following, we describe the entangled state
and its relation to Schmidt number in Sec.~\ref{sec:state_description}
and then show the fringe pattern simulations in Sec.~\ref{sec:simulation}.
Conclusions are given in Sec.~\ref{sec:conclusions}.

\section{Classical entangled state\label{sec:state_description}}

A beam traveling in the $Z$-direction can be expressed by the electric
field:

\begin{equation}
\mathbf{E}=E_{x}\hat{\mathbf{e}}_{x}+E_{y}\hat{\mathbf{e}}_{y}.\label{eq:elec_field}
\end{equation}
The \textbf{$\mathbf{E}$} field in Eq.~\eqref{eq:elec_field} exhibits
classical entanglement between two degrees of freedom in polarization
amplitudes and polarization directions,\textbf{ }where\textbf{ $\hat{\mathbf{e}}_{x}$}
and $\hat{\mathbf{e}}_{y}$ are the unit vectors in ``lab frame''
for two polarization directions, and $E_{x}$ and $E_{y}$ that are
vectors in ``function frame'' indicate the wave amplitudes~\citep{Qian}.
With the intensity $I=\left\langle E_{x}E_{x}+E_{y}E_{y}\right\rangle $,
the normalized electric field can be further expressed as:

\begin{equation}
\hat{\mathbf{e}}=\frac{\mathbf{E}}{\sqrt{I}}=\left(\cos\theta\Phi_{x}\hat{\mathbf{e}}_{x}+\sin\theta\Phi_{y}\hat{\mathbf{e}}_{y}\right),\label{eq:normalized_field}
\end{equation}
where $\Phi_{x}$ and $\Phi_{y}$ are the unit vectors in the function
frame referring to the relative amplitudes, which can be written as:

\begin{equation}
\Phi_{x}=\cos(kz-\phi_{x})
\end{equation}

\begin{equation}
\Phi_{y}=\cos(kz-\phi_{y}).
\end{equation}
Here, a phase difference between two functional vectors is $\Delta\phi=\phi_{y}-\phi_{x}$
, causing a nonzero cross correlation : 
\begin{equation}
\left(\Phi_{x},\Phi_{y}\right)=\frac{1}{\pi}\intop_{-\pi}^{\pi}\cos(kz-\phi_{x})\cos(kz-\phi_{y})\textrm{d}z=\cos\Delta\phi.
\end{equation}
Therefore, a new pair of orthogonal functions are chosen as $\Phi_{k}=\cos kz$,
$\Phi_{j}=\sin kz$ to guarantee $\left(\Phi_{k},\Phi_{j}\right)=0$.
Then,\textbf{ }the normalized field in Eq.~\eqref{eq:normalized_field}
can be rewritten by the new orthogonal vectors :

\begin{equation}
\hat{\mathbf{e}}=\left(\cos\phi_{x}\cos\theta\hat{\mathbf{e}}_{x}+\cos\phi_{y}\sin\theta\hat{\mathbf{e}}_{y}\right)\Phi_{k}+\left(\sin\phi_{x}\cos\theta\hat{\mathbf{e}}_{x}+\sin\phi_{y}\sin\theta\hat{\mathbf{e}}_{y}\right)\Phi_{j}.\label{eq:new basis normalized field}
\end{equation}
From Eq.~\eqref{eq:new basis normalized field}, the coefficient
matrix can be derived, serving as background in the Schmidt analysis:
\begin{equation}
C=\begin{pmatrix}\cos\phi_{x}\cos\theta & \sin\phi_{x}\cos\theta\\
\cos\phi_{y}\sin\theta & \sin\phi_{y}\sin\theta
\end{pmatrix}.\label{eq:coefficient matrix}
\end{equation}

The degree of entanglement decides how much separability the state
is, which can be evaluated by a Schmidt analysis in both quantum and
classical systems. According to Schmidt theorem~\citep{eberly_schmidt},
the Schmidt number $K$ can be calculated to define the degree of
entanglement precisely. In our case, the reduced density matrix of
the lab frame can be obtained from Eq.~\eqref{eq:new basis normalized field}
by tracing over the function frame:

\begin{equation}
\rho_{\textrm{lab}}=\begin{pmatrix}\cos^{2}\theta & \cos\Delta\phi\cos\theta\sin\theta\\
\cos\Delta\phi\cos\theta\sin\theta & \sin^{2}\theta
\end{pmatrix}.
\end{equation}
Then, the Schmidt number $K$ can be obtained by summing the squared
eigenvalues $\lambda_{s}^{2}$ of the reduced density matrix as weights,
taking the form~\citep{eberly_schmidt}:
\begin{equation}
K=\frac{1}{\sum_{s}\lambda_{s}^{2}}=\frac{1}{1-\frac{1}{2}\sin^{2}\Delta\phi\sin^{2}2\theta}=\frac{1}{1-2\textrm{det}(C)^{2}}.\label{eq:schmidt number}
\end{equation}
$K$ is a function of the determinant of the coefficient matrix, and
its value lies between 1 and 2 when two polarized dimensions are involved. 

When $K$ reaches the minimal value of unity, the electric field is
in a separable state where the two polarization components are linearly
polarized. Hence, the vectors appear in a dot product:
\begin{equation}
\hat{\mathbf{e}}=\left(\cos\theta\hat{\mathbf{e}}_{x}+\sin\theta\hat{\mathbf{e}}_{y}\right)\Phi_{k}.
\end{equation}
At the other extreme, when $K$ reaches the maximal value of two,
the electric field is in a maximally entangled state that can be expressed
as:

\begin{equation}
\hat{\mathbf{e}}=\frac{\sqrt{2}}{2}\hat{\mathbf{e}}_{x}\Phi_{k}+\frac{\sqrt{2}}{2}\hat{\mathbf{e}}_{y}\Phi_{j},
\end{equation}
referring to the case of circular polarization. The intermediate value
of $K$ between these two extremes represents a partially entangled
state called elliptic polarization. In general, the classical entanglement
between polarization amplitudes and polarization directions is intrinsically
related to the polarized states. 

\section{Simulation results\label{sec:simulation}}

In this section, the \textquotedbl phase method\textquotedbl{} and
\textquotedbl amplitude method\textquotedbl{} are proposed to estimate
the degree of entanglement in an incident laser beam based on the
fringe patterns of interference between four light sources. 

\subsection{Phase method}

\begin{figure}[h]
\includegraphics[width=10cm]{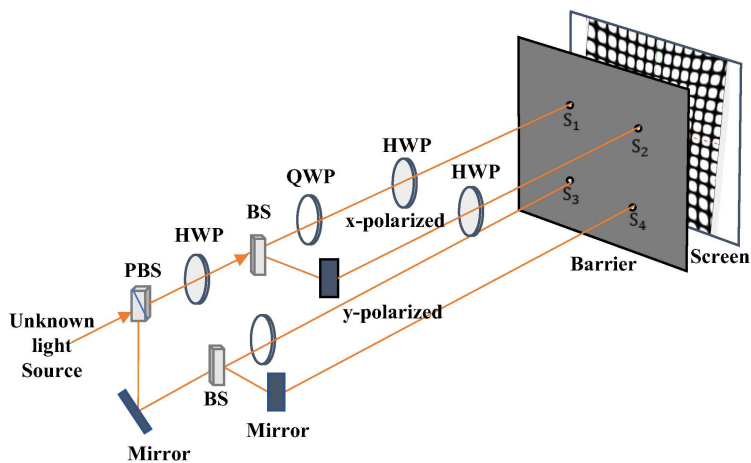}

\caption{Schematic diagram of the phase method. Firstly, the incident beam
impinges on a PBS to separate the horizontal and vertical polarization
beams. The split beams then go through two 50:50 BSs to create identical
copies separately. A QWP is placed at the transmitted output of the
beam splitter to cause a $\pi/2$-phase difference between beams in
the same polarization. Finally, the horizontally polarized beams are
converted into the vertical polarization direction by a rotating half-wave
plate at a $45^{\circ}$ orientation as inputs to the interference
sources.~\label{fig:phase_lightpath}}
\end{figure}

The patterns of interference between the four light sources are changed
due to the constructive and destructive effects at a different position.
We employ the schematic setup in Fig.~\ref{fig:phase_lightpath}.
An incident laser beam firstly impinges on a polarizing beam splitter
(PBS) to separate the horizontal and vertical polarization components,
where a $\pi/2$-phase shift is assigned to the reflected beam. Since
the mirror introduces another $\pi/2$-phase shift in the reflected
path, a half-wave plate (HWP) is applied in the transmitted path to
compensate for a total $\pi$-phase shift. Then, the split beams go
through two 50:50 beam splitters (BS) to create identical copies separately,
and a quarter-wave plate (QWP) is placed at the transmitted output
of the beam splitter to generate a $\pi/2$-phase difference between
beams in the same polarization. Finally, the horizontally polarized
beams are converted into a vertical polarization by a rotating half-wave
plate at a $45^{\circ}$orientation. 

The setup in Fig.~\ref{fig:phase_lightpath} decomposes the incident
beam into the four inputs of the interference experiment, which can
be expressed by the normalized input matrix $S$:

\begin{equation}
S=\begin{pmatrix}S_{11} & S_{12}\\
S_{21} & S_{22}
\end{pmatrix}=\begin{pmatrix}\cos\theta\cos(kz-\phi_{x}) & \cos\theta\sin(kz-\phi_{x})\\
\sin\theta\cos(kz-\phi_{y}) & \sin\theta\sin(kz-\phi_{y})
\end{pmatrix}=\begin{pmatrix}\cos\theta\Phi_{x} & \cos\theta\Phi_{x}^{\prime}\\
\sin\theta\Phi_{y} & \sin\theta\Phi_{y}^{\prime}
\end{pmatrix},\label{eq:inputmatrix1}
\end{equation}
where the matrix elements $S_{ij}$ refer to the input amplitude of
the four input sources in Fig.~\ref{fig:phase_lightpath}. The $S$
matrix contains all the polarization information of the original beam,
which can be extracted directly through individual photodetection
of each matrix element~\citep{Azzam}. For the proposal here, the
associated patterns projected on the screen in Fig.~\ref{fig:phase_lightpath}
is the main concern, which vary with the different values adopted
by the phase difference $\Delta\phi$ and the polarized angle $\theta$
and illustrate the degree of entanglement. The projected image reflects
the same determinant with the coefficient matrix~\eqref{eq:coefficient matrix},
$\textrm{det}(C)=\textrm{det}(S)$. Since the information of $\Delta\phi$
is stored in the phase of input fields, this method is the so-called
``phase method.''

\begin{figure}
\includegraphics[width=10cm]{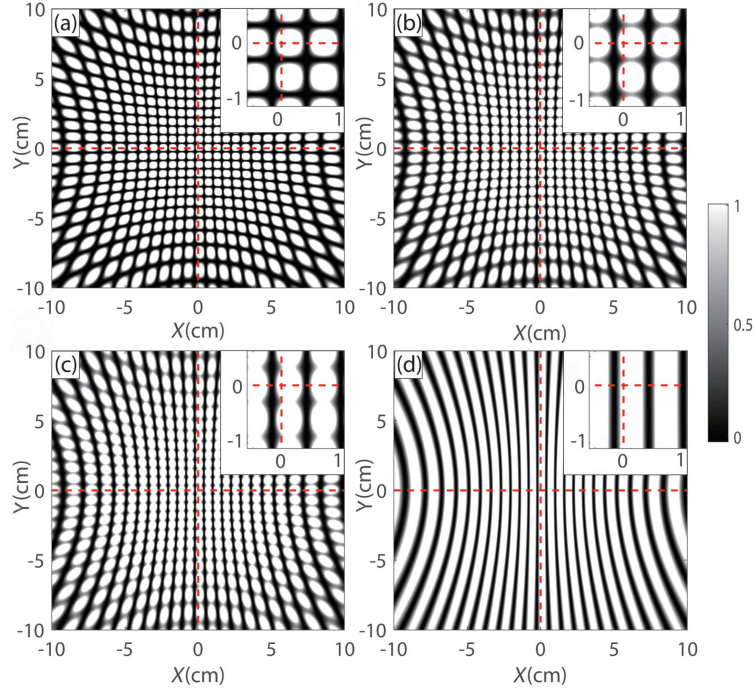}\caption{Interference patterns of the phase method are for the incident beams
in the separable states when $K=1$. The insets show the patterns
near the origin of the $XY$-plane in detail, indicating the fringes
are horizontally symmetric along the $X$-axis. These coefficients
in the normalized input matrix are: (a) $\Delta\phi=0$, $\theta=\pi/4$.
(b) $\Delta\phi=0$, $\theta=\pi/7$. (c) $\Delta\phi=0$, $\theta=\pi/10$.
(d) $\Delta\phi=\pi/2$, $\theta=\pi/2$. \label{fig:phase_pattern1}}
\end{figure}

\begin{figure}
\includegraphics[width=10cm]{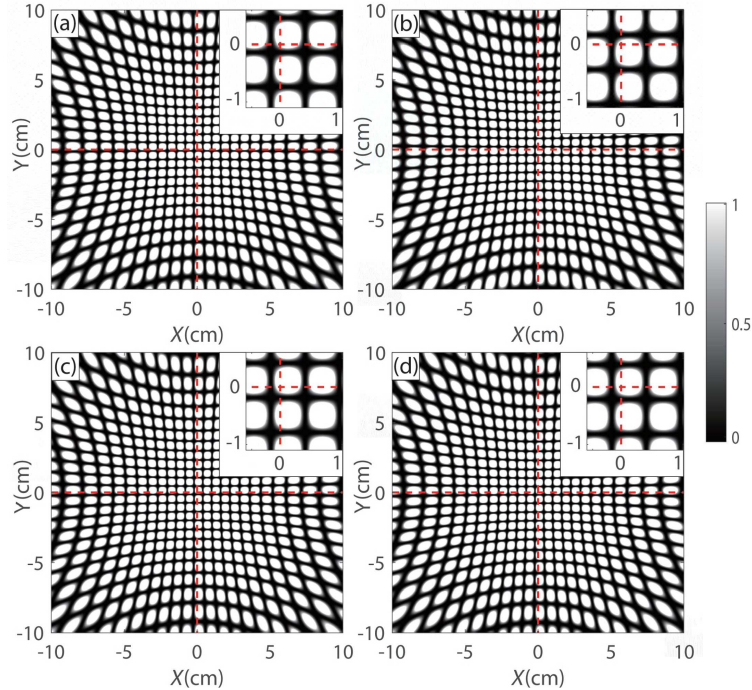}\caption{Interference patterns of the phase method in (a) and (b) are for the
incident beams in the maximally entangled states when $K=2$. The
patterns in (c) and (d) are for the intermediate scenarios when $1<K<2$.
The insets show the patterns near the origin of the $XY$-plane in
detail, indicating the changes in position of the horizontal symmetry
axis. These coefficients in the normalized input matrix are: (a) $K=2$:
$\Delta\phi=\pi/2$, $\theta=\pi/4$. (b) $K=2$: $\Delta\phi=-\pi/2$,
$\theta=\pi/4$. (c) $K=1.6$: $\Delta\phi=\pi/3$, $\theta=\pi/4$.
(d) $K=1.3$: $\Delta\phi=5\pi/21$, $\theta=\pi/4$.\label{fig:phase_pattern2}}
\end{figure}

A model of interference between four light sources is constructed
to simulate the interference patterns, which includes: the laser wavelength
of $600$~nm, the point-point gap of $d=10^{-5}$~m, the barrier-screen
distance of $L=0.3$~m, the screen area of $0.2\times0.2~\textrm{m}^{2}$.
The simulated patterns are shown in Figs.~\ref{fig:phase_pattern1}
and \ref{fig:phase_pattern2}, illustrating the interference intensity
distributions of incident beams in the separable states, the maximally
entangled states, and the intermediate scenarios.

The interference fringes for the scenarios when $K$ reaches its minimal
value of unity for the separable states are shown in Fig.~\ref{fig:phase_pattern1}.
The fringes have different patterns corresponding to distinct values
of the polarization angle $\theta$ for each subplot (a)-(c) when
$\Delta\phi=0$. These patterns refer to a zero phase difference between
the vertical pair of sources (either $S_{11}$ and $S_{21}$ or $S_{12}$
and $S_{22}$) from Eq.~\eqref{eq:inputmatrix1}. For the case shown
in subplot (d) where $\theta=\pi/2$, the fringes retain the same
pattern over different $\Delta\phi$ values because the effects from
sources $S_{11}$ and $S_{12}$ vanish, effectively making the case
a two-hole interference. It is worth noting that the patterns in all
cases of Fig.~\ref{fig:phase_pattern1} are symmetric about the $X$-axis,
signifying the horizontal symmetry in the fringes as a feature of
separable states.

In contrast, the maximally entangled states are obtained when $\theta=\pi/4$
while either $\Delta\phi=\pi/2$, as shown in Fig.~\ref{fig:phase_pattern2}(a),
or $\Delta\phi=-\pi/2$, as shown in Fig.~\ref{fig:phase_pattern2}(b).
Compared to the patterns of the separable states, the horizontal symmetry
axis can be moved either downward or upward from the $X$-axis for
a fixed distance, depending on the sign of $\Delta\phi$. Whereas
the vertical symmetry axis does not change due to a constant $\pi/2$-phase
difference between $S_{11}$ and $S_{12}$, $S_{21}$ and $S_{22}$.
Similarly, for the patterns of the intermediate scenarios in subplots
(c)-(d), the changes in the value of $\Delta\phi$ shift the horizontal
symmetry axis away from the $X$-axis along with the increase of $K$
while the vertical symmetry is retained. Shown in the insets of Fig.~\ref{fig:phase_pattern2},
the position of the horizontal symmetry axis follows the degree of
entanglement in reference to those of the limiting cases for maximally
entangled states and separable states.

Based on these analyses, the phase method can estimate the degree
of entanglement through the corresponding fringe patterns and their
horizontal symmetry axis. However, those changes in interference patterns
are not apparent to precisely distinguish the maximally entangled
state. A modified strategy called the ``amplitude method'' is introduced
with a phase analyzer and rotatable polarizers to improve detection
accuracy.

\subsection{Amplitude method}

The interference patterns vary directly with the polarization angle
and the phase differences among the four beam paths in the phase method.
Therefore, the phase method generates similar fringe patterns for
different states and the patterns are affected by the path drift to
some extent. In order to improve the phase method, we change the normalized
input matrix $S$ to the same form as the coefficient matrix in Eq.~\eqref{eq:coefficient matrix}:

\begin{equation}
S=\begin{pmatrix}\cos\phi_{x}\cos\theta & \sin\phi_{x}\cos\theta\\
\cos\phi_{y}\sin\theta & \sin\phi_{y}\sin\theta
\end{pmatrix}\Phi_{x}.\label{eq:input amplitude method}
\end{equation}
The new input matrix contains the phase information of $\phi_{x}$
and $\phi_{y}$ through the amplitudes of the four sources, and we
call the extraction of the phase information under such an $S$ matrix
the ``amplitude method.'' In the experimental setup, the interference
fringes rely on the phase differences $\Delta\phi$ related to the
$S$ matrix elements through the formula:
\begin{equation}
\sin\phi_{y}\cos\phi_{x}-\sin\phi_{x}\cos\phi_{y}=\text{\ensuremath{\sin}}\Delta\phi
\end{equation}
and experimentally extracted with the assistance of a phase analyzer.

\begin{figure}[h]
\includegraphics[width=10cm]{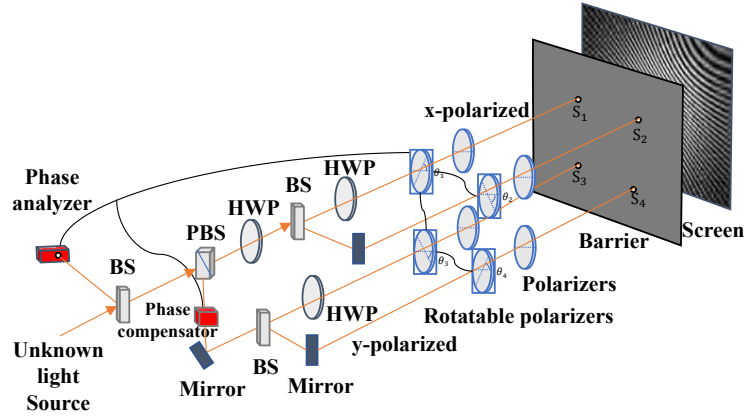}\caption{Schematic diagram of the amplitude method. The incident beam firstly
goes through a 50:50 BS whose reflected beam is directed to the phase
analyzer. The transmitted beam is divided into the horizontal and
vertical polarization beams by passing through a PBS. Then, the split
beams are decomposed by 50:50 BSs and directed to the rotatable polarizer
array. The amplitude coefficients of the phases $\phi_{x}$ and $\phi_{y}$
are realized by the angle of rotation $\theta_{n}$ in the rotatable
polarizers.\label{fig:amp_lightpath}}
\end{figure}

The schematic for the proposed setup is shown in Fig.~\ref{fig:amp_lightpath}.
The incident beam is first sent to a 50:50 BS whose reflected part
is directed to a phase analyzer. The key phase difference $\Delta\phi$
is extracted at this early stage of the light path, mitigating the
adverse effect of path drift. The transmitted beam is then separated
into the horizontal and vertical polarization beams by passing through
a PBS. An HWP and a phase compensator are then placed at the output
of the PBS to compensate for a $\pi$-phase shift introduced by the
reflection in the mirror and PBS, and its $\Delta\phi$ phase difference.
After the phase compensation, the split beams are decomposed by a
50:50 BS, where another HWP is placed at the transmitted output to
cancel the reflected phase shift. Consequently, the four beams are
changed into a horizontal polarization by the polarizer array that
consists of rotatable polarizers and $90^{\circ}$ polarizers. Meanwhile,
the amplitude coefficients of the phases $\phi_{x}$ and $\phi_{y}$
are realized by the angle of rotation $\theta_{n}$ in the rotatable
polarizers due to Malus' law, i.e. 

\begin{equation}
\cos\phi_{x}=\sin\theta_{1}\cos\theta_{1}.
\end{equation}
This method finally decomposes the incident beam into four coherent
sources for interferometry as: $S_{11}=\cos\phi_{x}\cos\theta\Phi_{x}$;
$S_{12}=\sin\phi_{x}\cos\theta\Phi_{x}$; $S_{21}=\cos\phi_{y}\sin\theta\Phi_{x}$;
$S_{22}=\sin\phi_{y}\sin\theta\Phi_{x}$.

\begin{figure}
\includegraphics[width=10cm]{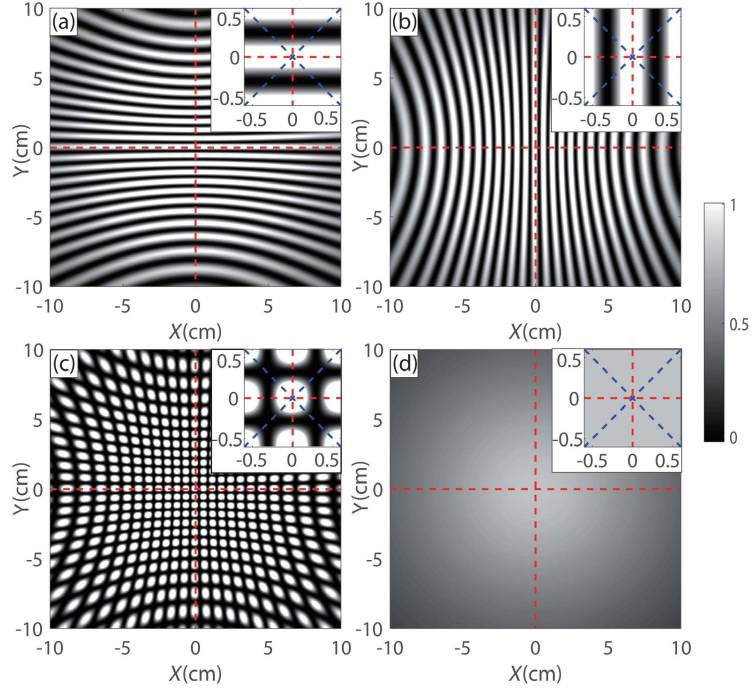}\caption{Interference patterns of the amplitude method are for the incident
beams in the separable states when $K=1$. The insets magnify the
central unit pattern, indicating that the fringes are symmetric along
the \ensuremath{X}- or \ensuremath{Y}- direction. These coefficients
in the normalized field matrix are: (a) $\sin\phi_{x}=\sin\phi_{y}=0$
and $\theta=\pi/4$; (b) $\sin\phi_{x}=\sin\phi_{y}=\cos\phi_{x}=\cos\phi_{y}=\sqrt{2}/2$
and $\theta=\pi/2$; (c) $\sin\phi_{x}=\sin\phi_{y}=\cos\phi_{x}=\cos\phi_{y}=\sqrt{2}/2$
and $\theta=\pi/4$; (d) $\cos\phi_{x}=\cos\phi_{y}=1$ and $\theta=\pi/2$.~\label{fig:amp_pattern1}}
\end{figure}

Another model is established with the same setup in the first method,
which includes: the laser wavelength of $600$~nm, the point-point
gap of $d=10^{-5}$~m, the barrier-screen distance of $L=0.3$~m,
the screen area of $0.2\times0.2~\textrm{m}^{2}$. In the simulation,
incident beams are chosen as the separable states, the maximally entangled
states, and the intermediate scenarios.

\begin{figure}
\includegraphics[width=10cm]{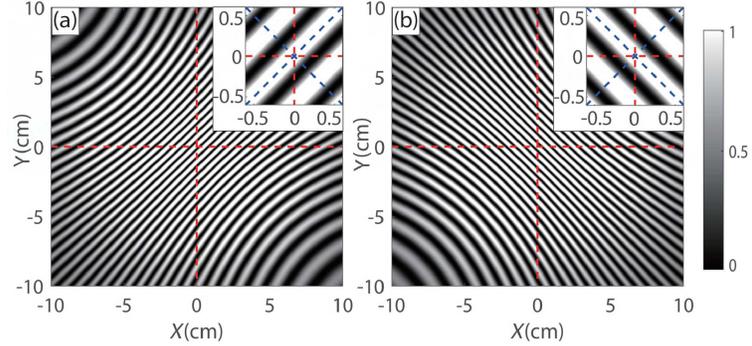}\caption{Interference patterns of the amplitude method are for the incident
beams in the maximally entangled states when $K=2$. The insets magnify
the central unit pattern, indicating that the fringes are symmetric
along the axis at $45^{\circ}$ or $135^{\circ}$ direction. These
coefficients in the normalized field matrix are: (a) $\sin\phi_{y}=\cos\phi_{x}=1$
and $\theta=\pi/4$. (b) $\sin\phi_{x}=\cos\phi_{y}=1$ and $\theta=\pi/4$.\label{fig:amp_pattern2}}
\end{figure}

The interference fringes of the separable states are shown in Fig.~\ref{fig:amp_pattern1}.
The subplots (a)-(d) illustrate different patterns, where the fringes
are all symmetric along the $X$- or $Y$-axis or simultaneously.
This symmetry depends on the zero determinant in Eq.~\eqref{eq:input amplitude method}
when $K=1$. For example, the subplot (a) indicates the fringes distributed
along the $Y$-axis when a vertical pair of sources $S_{11}$ and
$S_{21}$ interfere since the coefficients ($\sin\phi_{x}=\sin\phi_{y}=0$
and $\theta=\pi/4$) generate the input amplitude in Eq.~\eqref{eq:input amplitude method}
as $S_{11}=S_{21}=\sqrt{2}/2$ and $S_{12}=S_{22}=0$. In addition,
the subplot (d) demonstrates the scenario that only the source $S_{11}$
works when $\cos\phi_{x}=\cos\phi_{y}=1$ and $\theta=\pi/2$. For
the case of the maximally entangled states, $K$ reaches its maximal
value of two when $\theta=\pi/4$ while either $\sin\phi_{y}=\cos\phi_{x}=1$,
or $\sin\phi_{y}=\cos\phi_{x}=1$, which refers to the diagonal pair
of sources (either $S_{11}$ and $S_{22}$ or $S_{12}$ and $S_{21}$)
interfere. The dark and bright fringes of the maximally entangled
states are symmetric along the diagonal axis at $45^{\circ}$ or $135^{\circ}$
direction, while they are randomly distributed in the \ensuremath{X}-
and \ensuremath{Y}- axes. Therefore, in Fig.~\ref{fig:amp_pattern2},
the fringe patterns are different from those in the separable states,
which can easily distinguish these two cases.

\begin{figure}
\includegraphics[width=10cm]{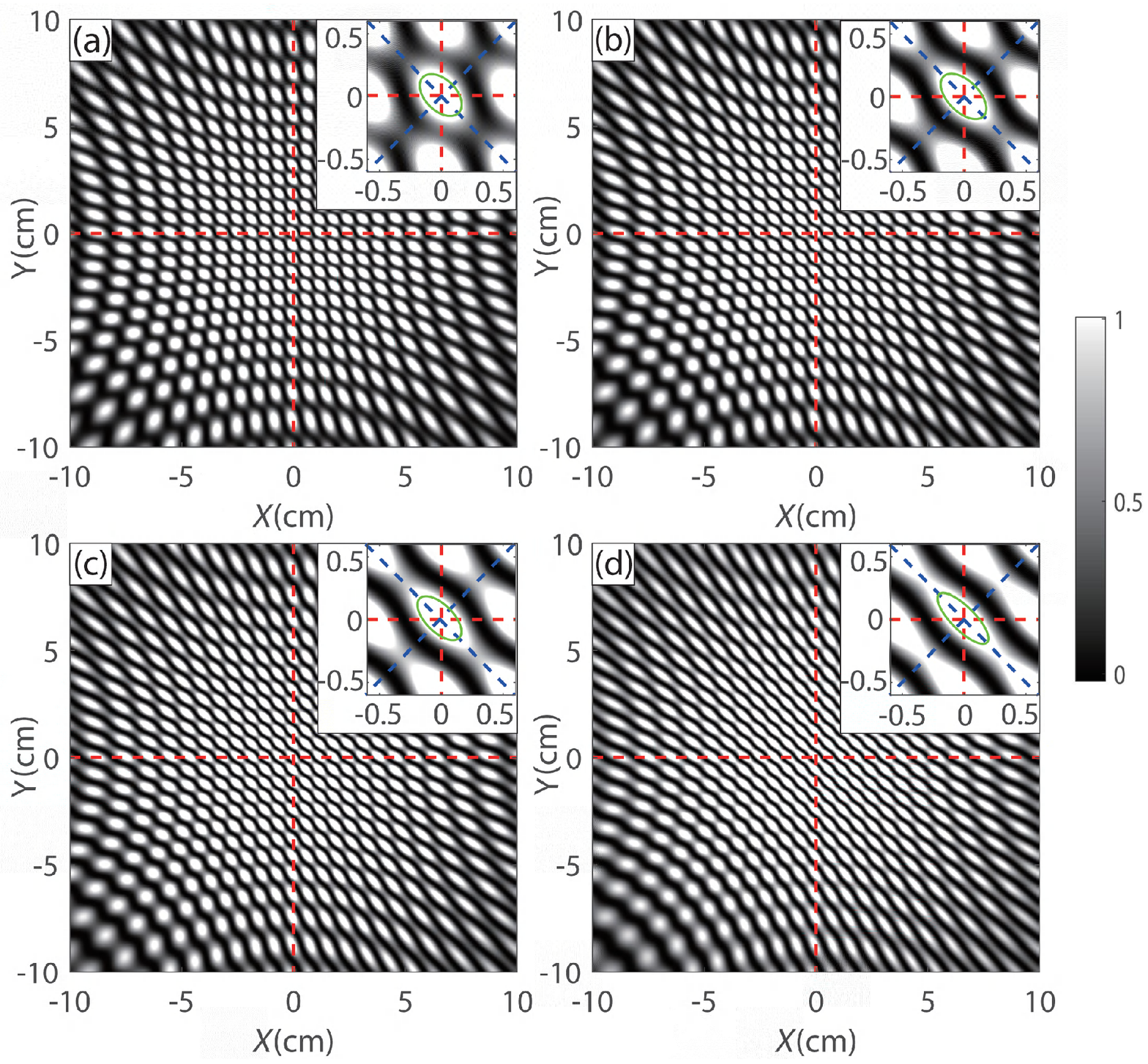}\caption{Interference patterns of the amplitude method are for the incident
beams in the intermediate scenarios when $1<K<2$. These coefficients
in the normalized field matrix are: (a) $K=1.2$: $\cos\phi_{x}=0.888,\sin\phi_{x}=0.459,\cos\phi_{y}=0.459,\sin\phi_{y}=0.888$
and $\theta=\pi/4$; (b) $K=1.4$: $\cos\phi_{x}=0.936,\sin\phi_{x}=0.349,\cos\phi_{y}=0.349,\sin\phi_{y}=0.936$
and $\theta=\pi/4$; (c) $K=1.6$: $\cos\phi_{x}=0.965,\sin\phi_{x}=0.258,\cos\phi_{y}=0.258,\sin\phi_{y}=0.965$
and $\theta=\pi/4$; (d) $K=1.8$: $\cos\phi_{x}=0.985,\sin\phi_{x}=0.169,\cos\phi_{y}=0.169,\sin\phi_{y}=0.985$
and $\theta=\pi/4$. Contours (green line) at the same light intensity
are plotted in the insets.\label{fig:amp_pattern3}}
\end{figure}

The fringes for the intermediate scenarios are illustrated in Fig.~\ref{fig:amp_pattern3},
where the patterns could be viewed as transiting from those for separable
states to those for maximally entangled states. In particular, the
symmetry axes of fringes are rotating toward $45^{\circ}$ or $135^{\circ}$
directions from the \ensuremath{X}- or \ensuremath{Y}-axis along with
the increase of $K$. Based on this feature, the rotation of symmetry
axes can visually distinguish the intermediate scenarios from the
separable states. For example, Fig.~\ref{fig:amp_pattern3}(a) demonstrates
that an intermediate state with $K$ closed to unity is distinct from
the separable state shown in Fig.~\ref{fig:amp_pattern1}(c). Then
along with the further increase of $K$, as shown in Fig.~\ref{fig:amp_pattern3}(b)-(d),
the patterns retain the same symmetry axis at $45^{\circ}$ or $135^{\circ}$
direction while the bright fringe is further stretched along the $135^{\circ}$
axis. The contour levels at the same light intensity, showing quasi-elliptical
patterns, are added in the insets of Fig.~\ref{fig:amp_pattern3}
to assist the quantification of such changes.

\section{Entanglement-fringe pattern relationship}

\begin{figure}
\includegraphics[width=10cm]{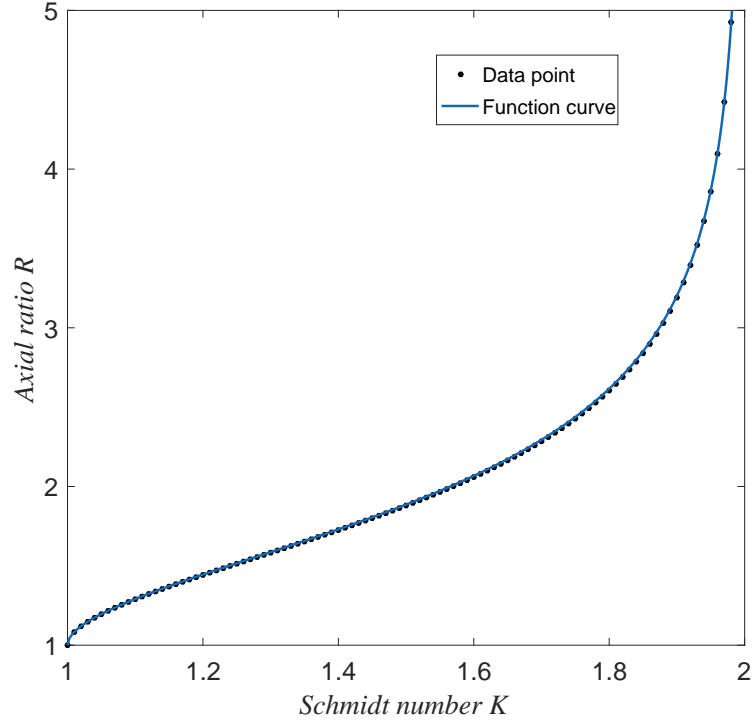}\caption{The axial ratio $R$ as a function of Schmidt number is plotted from
both the numerical simulations (black dots) and the analytical formula
derived as Eq.~\eqref{eq:axial ratio} (blue curve). The two approaches
agree well in the plots and the unique one-one correspondence between
the fringe pattern and the entanglement degree is established.~\label{fig:The-length-ratio}}
\end{figure}

The stretching along the $135^{\circ}$ axis is quantified by the
major-to-minor axis ratio $R$ of the contours and Fig.~\ref{fig:The-length-ratio}
shows the variation of this ratio at various values of Schmidt number
$K$, where the data points are extracted from the numerical analysis
from the last section. Though not a linear function of $K$, the ratio
has a one-one (bijective) correspondence with $K$. In other words,
through observing this characteristic of axis ratio in the fringe
pattern of a classically entangled beam, one can obtain a unique Schmidt
number for entanglement degree.

To show the one-one correspondence exist between $K$ and $R$ analytically,
we consider that the intensity of the image distributed on the screen
is $I=|E|^{2}$ where the electric field 
\begin{equation}
E=\sum_{i,j=1}^{2}\frac{E_{0}\exp\left\{ ik\sqrt{[x+(-1)^{j}a]^{2}+[y+(-1)^{i}a]^{2}+b^{2}}\right\} }{\sqrt{[x+(-1)^{j}a]^{2}+[y+(-1)^{i}a]^{2}+b^{2}}}S_{ij}\label{eq:E_field}
\end{equation}
is the superposition of the elements of the source matrix $S$ of
Eq.~\eqref{eq:input amplitude method} weighted by the phases and
scales contributed by the source-to-screen distance. In the equation,
$x$ and $y$ are the horizontal and vertical distance to the screen
center of the image; $a$ is half of the inter-hole distance of the
four holes shown in Fig.~\ref{fig:amp_lightpath}; and $b$ is the
perpendicular screen-to-image distance.

For the fringe image, we consider the central pattern where $x,y\ll b$
and $x,y\gg a$, such that the scale weights can be approximated by
$1/b$ (zeroth power approximation) while the phase weights by $k\sqrt{x^{2}+y^{2}+b^{2}+2a^{2}}+k(\pm x\pm y)a/b$
(first-power approximation). For the typical interferences shown in
Figs.~\ref{fig:amp_pattern1}, \ref{fig:amp_pattern2}, and \ref{fig:amp_pattern3},
we consider only $\theta=\pi/4$ to simplify the calculation without
loss of generality. These considerations give rise to the intensity
formula
\begin{equation}
I=\frac{E_{0}^{2}}{b^{2}}\left\{ 1\pm\sin\Delta\phi+2\cos\Delta\phi\cos t(z)+\left(1\mp\sin\Delta\phi\right)\cos^{2}t(z)\right\} \label{eq:intensity}
\end{equation}
on the diagonal $x\pm y=0$ axes, where $z=\sqrt{x^{2}+y^{2}}=\sqrt{2}x$
and
\begin{equation}
t(z)=\frac{\sqrt{2}akz}{\sqrt{z^{2}+b^{2}+2a^{2}}}.\label{eq:t_z}
\end{equation}
The intensity \eqref{eq:intensity} is upper bounded by $4E_{0}^{2}/b^{2}$.
We can take half of this value as an arbitrary contour level to delineate
the quasi-elliptical pattern. The constant value corresponds to a
quadratic equation of $\cos t(z)$, the roots of which associate with
the intersection points of the contour and the diagonal axes. Fixing
$t$ in Eq.~\eqref{eq:t_z}, the equation $z=t\sqrt{(b^{2}+2a^{2})/(2a^{2}k^{2}-t^{2})}$
reflects the lengths of the major or minor axis of the quasi-elliptical
pattern.

To establish the relation with the Schmidt number $K$, we solve the
quadratic equation to find (taking only the root with positive determinant)
\begin{equation}
\cos t(z)=\frac{-\cos\Delta\phi+\sqrt{2(1\mp\sin\Delta\phi)}}{1\mp\sin\Delta\phi}.
\end{equation}
According to Eq.~\eqref{eq:schmidt number}, $\sin\Delta\phi=\sqrt{2-2/K}$
and thus $\cos\Delta\phi=\sqrt{2/K-1}$. Considering the approximation
$z,a\ll b$ as before, $t(z)$ becomes linear in $z$ and we find
directly the ratio of the major to the minor axis to be
\begin{equation}
R(K^{\prime})=\frac{\cos^{-1}\left(\left[\sqrt{2}-\sqrt{1+K^{\prime}}\right]/\sqrt{1-K^{\prime}}\right)}{\cos^{-1}\left(\left[\sqrt{2}-\sqrt{1-K^{\prime}}\right]/\sqrt{1+K^{\prime}}\right)},\label{eq:axial ratio}
\end{equation}
where $K^{\prime}=\sqrt{2-2/K}$. The analytical expression \eqref{eq:axial ratio}
is plotted as the blue curve in Fig.~\ref{fig:The-length-ratio},
which fits well with the data points extracted from numerical simulations.
Therefore, one can quantitatively extract a unique Schmidt number
from the corresponding interference fringes by measuring the axial
ratio and the discernable angles. The one-one relationship between
fringe patterns and the entanglement degree in a classical beam is
established.

\section{Conclusions and discussions\label{sec:conclusions}}

We employed the concept of classical entanglement between the polarization
amplitude and the polarization direction of an optical beam to study
the relation between the degree of entanglement and the interference
fringe pattern hidden in this beam. Two methods (we named them the
phase method and the amplitude method, respectively) based on optical
manipulations to separate and interfere the polarization components
of the beam are proposed to generate the fringe pattern. Both methods
demonstrate differing patterns corresponding to distinct entangled
states measured by Schmidt number. The amplitude method improves over
the phase method in terms of the easiness in differentiating the patterns,
where the maximally entangled states for instance correspond to fringes
exactly symmetric about the diagonal axes. Along these lines, the
methods proposed would be beneficial to realizing quantum algorithms
using classically entangled states.

We note that in actual experimental setups, the fringe patterns might
be smeared by speckle effects during projection. Nonetheless, statistical
analysis such as linear regression can be used as image processing
techiques to extract key features like the axial ratio despite the
distorted projection image. Some experiments~\citep{Emile14} have
already employed such techiques to analyze interference patterns.
Also, though the amplitude method has improved on the phase method
in regards to the reduction of path drift, drifting can still exist.
In this case, active mechanisms such as those relying on position-sensitive
detection~\citep{Hu} can be introduced along the light paths to
mitigate the effect.

\section*{Acknowledgments}

H.I. thanks the support by FDCT of Macau under Grant 0130/2019/A3
and by University of Macau under MYRG2018-00088-IAPME.

\end{document}